\documentclass[twocolumn,showpacs,preprintnumbers,amsmath,amssymb]{revtex4}
\usepackage[dvips]{graphicx} 
\usepackage[english]{babel}

\begin{document}

\preprint{IFF-RCA-4-12}
\title{Entanglement arrow of time in the multiverse}
\author{Salvador Robles-P\'{e}rez}
\affiliation{Colina de los Chopos, Centro de F\'{\i}sica ``Miguel Catal\'{a}n'',
Instituto de Matem\'{a}ticas y F\'{\i}sica Fundamental,
Consejo Superior de Investigaciones Cient\'{\i}ficas, Serrano 121, 28006
Madrid (SPAIN) and \\ Estaci\'{o}n Ecol\'{o}gica de Biocosmolog\'{\i}a, Pedro de Alvarado, 14, 06411-Medell\'{\i}n, (SPAIN).}
\date{\today}

\begin{abstract}
In this paper it is presented the model of a multiverse made up of entangled pairs of universes. The arrow of time obtained from the principles of thermodynamics and the arrow of time given by the thermodynamics of entanglement for single universes are analyzed. The latter requires that the single universes expand once they have crossed the quantum barrier at the Euclidean regime. The possible relationship with respect to the grow of local structures in a single universe is also discussed. 
\end{abstract}

\pacs{98.80.Qc, 03.65.Ud}
\maketitle

\section{Introduction}

One of the most exciting and difficult challenges of physics consists of providing a satisfactory explanation of the arrow(s) of time. The difficulties are at least two fold. On the one hand, it must cover all the different arrows of time that are customary considered \cite{Hawking1985}, i.e. the thermodynamic, the cosmological, the electromagnetic, and the psychological arrows of time. On the other hand, the origin of the arrow of time can be tracked to the very early conditions of the universe, a question which is far from being solved.

Proof of the fascinating character of this problem is the number of papers, conferences and publications that can be found in the literature (for extended reviews, see Refs. \cite{ Davies1974, Zeh1989, Halliwell1994}, and references therein). Even restricting ourselves to cosmological grounds, the arrow of time has been studied in relation to the large scale expansion of the universe \cite{Halliwell1994, Kiefer1995},  to the boundary conditions of the universe \cite{Hawking1985, Wald2005}, in string theory and brane worlds \cite{McInnes2007, Gorsky2007, Bousso2011}, in inflationary models \cite{Goldwirth1991, Albrecht2003, Carroll2004}, and in quantum cosmology \cite{Hawking1985, PFGD1994, Castagnino2002, Kiefer2005, Alfinito2007}.

Needless to say, the vast majority of the references deal with the arrow of time in the context of a single universe (for an interesting exception, see Ref. \cite{Bertolami2008}). The multiverse, in its wide variety of interpretations (see Ref. \cite{Carr2007}), constitutes a new cosmological paradigm that adds novel features to the universe and, thus, it opens the door to new ways of facing up the problem of the arrow of time. First of all, in the context of a physical multiverse \cite{PFGD2012} the universe is no longer an isolated system and it must be  treated as an open system which  interacts with the rest of universes of the multiverse. In particular, classical and quantum correlations can be present in the state of the quantum multiverse. The former would be induced by the existence of wormholes that would crop up and join different regions of two or more universes \cite{PFGD2012}, and the latter would also exist provided that entangled and other non-classical states can be present in the quantum multiverse \cite{RP2010, RP2012}.

Furthermore, the boundary conditions of the whole multiverse can be quite different from those imposed on single universes. They may be such that cosmological quantum effects in the form of entanglement between universes would survive the Euclidean barrier and still be present along the whole evolution of a large parent universe \cite{RP2012}. In quantum information theory, an arrow of time in terms of the entropy of entanglement between two correlated systems has already been given in Ref. \cite{Jennings2010}. Then, the entropy of entanglement between two universes of the multiverse can also supply us with an arrow of time for single universes.

It is not clear yet the relationship between the classical and the quantum arrows of time as well as the relationship between classical thermodynamics and the thermodynamics of entanglement (for recent developments, see Refs. \cite{Vedral2002, Brandao2008}). An extension of the classical laws of thermodynamics would be expected that account for both classical and entanglement thermodynamics provided that quantum theory is considered a more general framework from which the classical one is recovered as a particular limiting case. Then, the classical and the quantum entropic processes in the multiverse would be related, too.

The formalism of the quantum information theory can be applied in the quantum multiverse \cite{Deutsch2002}. In a third quantization formalism \cite{Strominger1990}, a well-defined Fock space can be posed for the quantum state of a multiverse made up of universes with high order of symmetry \cite{RP2010}. The universes can be created in correlated pairs as a consequence of the boundary conditions of the multiverse  \cite{RP2010, RP2012} and, then, the quantum correlations between the universes and the thermodynamical properties of entanglement can be computed by using the usual techniques of quantum information theory, in the conceptual context though of a multiverse  \cite{RP2012}.

In this paper, we thus present  a model of the quantum multiverse in which an arrow of time can be defined both for single universes and for the multiverse itself, within the framework of a quantum information theory of the whole multiverse. We shall take the point of view of Ref. \cite{Hawking1985}, where the electromagnetic and the psychological arrows of time are consider consequences of the (classical) thermodynamic arrow of time. We would like to notice, however, that the latter might also be related to the quantum thermodynamical arrow of time provided that the brain might be considered  as a quantum system, which is currently a matter of investigation (for a discussion on the subject, see Ref. \cite{Litt2006} and references therein). Nevertheless, we shall only deal with the thermodynamical, quantum and classical, arrow of time and its relationship with the large scale expansion of single universes. 

In Sec. II, we fix the model and  impose appropriate boundary conditions to obtain the state of the whole multiverse, in which the universes are  created in correlated pairs. The entropy of entanglement is considered in Sec. III as an arrow of time for each single universe and the second(s) principles of thermodynamics, both classical and quantum, are analyzed in the context of the model as well as their relationship with the large scale expansion of the single universes. In Sec. IV we draw some conclusions and make further comments.

\section{Entangled pairs of universes in the quantum multiverse}

Let us consider a multiverse made up of closed Friedmann-Robertson-Walker universes endorsed with a cosmological constant, $\Lambda$, and a massless scalar field, $\varphi$. Cosmic quantum effects in the form of entanglement between the quantum states of two universes can survive the quantum regime and be present in the Lorentzian regime along the whole history of each single universe \cite{RP2010, RP2012}. Therefore, the homogeneity and isotropic assumptions are justified provided that we are dealing with a multiverse made up of large universes, for instance, of a length scale of order of the Hubble length of our universe. 

For such a model of universes, the Wheeler-DeWitt equation, with appropriate units and rescaling the scalar field to absorb unimportant constants, can be written  as \cite{Kiefer2007}
\begin{equation}\label{WDW}
(\hbar^2 a^2 \partial^2_{aa} + \Lambda a^{6} - a^4 - \hbar^2 \partial^2_{\varphi \varphi} ) \phi(a, \varphi) = 0 ,
\end{equation}
where $\phi(a,\varphi)$ is the wave function of the universe and, for simplicity, the factor ordering term has been disregarded  \footnote{The factor ordering term can be seen in Eq. (\ref{HO}) as a mass term that depends on the scale factor, which is the time variable in the third quantization formalism of the multiverse. The states of a harmonic oscillator with a \emph{time}-dependent mass increase the rate of squeezing of the state of the harmonic oscillator. They do not  introduce therefore any qualitative difference with respect to the case considered in Eq. (\ref{WDW}) in which squeezing effects are already present, as it will be shown in the rest of the section.} in Eq. (\ref{WDW}). In the third quantization formalism \cite{Strominger1990, RP2010, RP2012}, the wave function $\phi(a,\varphi)$ is promoted to an operator that can be decomposed in normal modes as 
\begin{equation}\label{decomposition}
\hat{\phi}(a,\varphi) = \int dk \; e^{i k \varphi} A_k^*(a) \hat{c}_k + e^{- i k\varphi} A_k(a) \hat{c}_k^\dag ,
\end{equation}
where, $\hat{c}_k^\dag \equiv \sqrt{\frac{\omega_0}{2 \hbar}} (\phi_k -\frac{i}{\omega_0}p_{\phi_k})$ and $\hat{c}_k \equiv \sqrt{\frac{\omega_0}{2 \hbar}} (\phi_k +\frac{i}{\omega_0}p_{\phi_k})$, are constant operators that represent the creation and annihilation of universes with an energy density which is proportional to $\omega_0$. The probability amplitudes, $A_k(a)$, satisfy the equation of a harmonic oscillator,
\begin{equation}\label{HO}
\ddot{A}_k(a) + \omega_k^2(a) A_k(a) = 0 ,
\end{equation}
with a \emph{time} dependent frequency given by, $ \omega_k^2(a) = \frac{\Lambda a^4 - a^2}{\hbar} +\frac{ k^2}{a^2}$, where the scale factor formally plays the role of the time variable, and  $\ddot{A}_k \equiv \frac{\partial^2 A_k}{\partial a^2}$ in Eq. (\ref{HO}).

The kind of universes created and annihilated by $\hat{c}_k^\dag$ and $\hat{c}_k$, respectively, depends on the boundary conditions which are imposed on the solutions of Eq. (\ref{HO}), i.e. it depends on the boundary conditions imposed on single universes. If the tunneling boundary condition is chosen, the only modes that survive the quantum barrier are the expanding branches of the universe \cite{Vilenkin1982, Vilenkin1984, Vilenkin1986}. Then, in the Lorentzian region, with $a\gg\frac{1}{\sqrt{\Lambda}}$, the probability amplitudes take the semiclassical form, $A_k \propto \frac{1}{a} e^{-\frac{i}{\hbar}S_c(a)}$, where $S_c(a) \approx \sqrt{\Lambda}a^3$, is the classical action. If we otherwise impose the no-boundary condition \cite{Hawking1982, Hartle1983}, the universes created and annihilated by $\hat{c}_k^\dag$ and $\hat{c}_k$, respectively, correspond to linear combinations of expanding and contracting branches of the universe and, in the case being considered, the probability amplitudes take the assymptotic form, $A_k \propto \frac{1}{a} \cos(\frac{S_c}{\hbar}+ \frac{i \pi k}{6})$, in the semiclassical limit.

However, the creation and annihilation operators $\hat{c}_k^\dag$ and $\hat{c}_k$ cannot properly be interpreted as the creation and annihilation operator of universes in the multiverse because, then, the number of universes of the multiverse, given by the eigenvalue of the operator $\hat{N}_k\equiv \hat{c}_k^\dag \hat{c}_k$, would depend on the value of the scale factor of each single universe. In order to see that, let us notice that the Hamiltonian of the harmonic oscillator that  quantum mechanically represents the dynamic evolution (\ref{HO}) of the multiverse  reads,
\begin{equation}\label{Ham1}
\hat{H} = \frac{1}{2} \hat{p}_\phi^2 + \frac{\omega_k^2(a)}{2} \hat{\phi}^2 ,
\end{equation}
and $[\hat{H}, \hat{N}_k]\neq 0$, for any mode $k$. It is not expected that the number of universes in the whole multiverse depends on the value of the scale factor of a particular single universe. Then, the boundary condition of the multiverse that \cite{RP2012} the number of universes does not depend on the value of the scale factor fixes the representation that has to be used. This is given by the Lewis representation \cite{Lewis1969, RP2010}, defined in terms of the following creation and annihilation operators,
\begin{eqnarray}\label{b1}
\hat{b}(a) &=& \sqrt{\frac{1}{2 \hbar}} \left( \frac{\hat{\phi}}{R} + i (R \hat{p}_{\phi} - \dot{R} \hat{\phi}) \right) , \\ \label{b2}
\hat{b}^\dag(a) &=& \sqrt{\frac{1}{2 \hbar}} \left( \frac{\hat{\phi}}{R} - i (R \hat{p}_{\phi} - \dot{R} \hat{\phi}) \right) ,
\end{eqnarray}
where, $R \equiv R_k(a) = \sqrt{\phi_1^2(a) + \phi_2^2(a)}$, with $\phi_1$ and $\phi_2$ being, for each mode $k$, two linearly independent solutions of Eq. (\ref{HO})  that make real the value of the function $R(a)$ \footnote{This is a necessary condition for the invariant operator, $\hat{I} \equiv \hat{b}^\dag(a) \hat{b}(a) + \frac{1}{2}$, to be Hermitian.}. The Lewis representation given by the operators (\ref{b1}-\ref{b2}) conserves the number of universes in the multiverse, i.e. $b^\dag b |N,a\rangle = N |N,a\rangle$, with $N\neq N(a)$. Thus, it can properly interpreted as representing the number of universes in the multiverse.

In terms of the creation and annihilation operators of the Lewis representation, the decomposition of the wave function (\ref{decomposition}) turns out to be
\begin{equation}\label{decomposition2}
\phi(a,\varphi) = \int dk \;  B_k(\varphi, a) \hat{b}_k(a) + B_k^*(\varphi,a) \hat{b}_k^\dag(a) ,
\end{equation}
with, $B_k(\varphi, a) = e^{i k\varphi} A_k(a) \mu^* - e^{-i k \varphi} A_k^*(a) \nu^*$, where, $\mu\equiv \mu(a)$ and $\nu\equiv \nu(a)$, are the squeezing parameters that relate the Lewis representation to the representation given by the constant operators $\hat{c}_k$ and $\hat{c}_k^\dag$, i.e. 
\begin{eqnarray}
\hat{b}_k &=& \mu \, \hat{c}_k + \nu \,  \hat{c}_k^\dag , \\
\hat{b}_k^\dag &=& \mu^* \,  \hat{c}_k^\dag + \nu^* \,  \hat{c}_k ,
\end{eqnarray}
with,
\begin{eqnarray}
\mu &=& \frac{1}{2\sqrt{\omega_0}} (\frac{1}{R} + \omega_0 R - i \dot{R}) , \\
\nu &=& \frac{1}{2\sqrt{\omega_0}} (\frac{1}{R} - \omega_0 R - i \dot{R}) ,
\end{eqnarray}
and $|\mu|^2 - |\nu|^2 = 1$. Thus, the Lewis operators given by Eqs. (\ref{b1}-\ref{b2}) do not represent the creation and annihilation of either an expanding or an equally probable combination of expanding and contracting branches of the universe but a scale factor dependent linear combination of expanding and contracting branches of the universe.

Therefore, the boundary condition of the multiverse that the number of universes does not depend on the value of the scale factor restricts the boundary conditions that can be imposed on single universes.

Moreover, the boundary condition of the quantum state of the multiverse implies that the universes have to be created (or annihilated) in correlated pairs. In terms of the Lewis representation, the Hamiltonian (\ref{Ham1}) turns out to be
\begin{equation}\label{H2}
\hat{H} = \hbar \left( \beta_- \hat{b}^2 + \beta_+ (\hat{b}^\dag)^2 + \beta_0 (\hat{b}^\dag \hat{b} + \frac{1}{2}) \right) ,
\end{equation}
where, $\beta_\pm \equiv \beta_\pm(a)$ and $\beta_0 \equiv \beta_0(a)$, are two functions that depend non-trivially on $R(a)$ \cite{RP2010, RP2012}. The structure of the Hamiltonian (\ref{H2}), with the quadratic terms in $\hat{b}$ and $\hat{b}^\dag$, allows us to interpret that the universes are created in the quantum multiverse in correlated pairs. Then, the entropy of entanglement between two entangled universes can be taken as the arrow of time for each single universe, as we shall see in the next section.

\section{Arrow of time for single universes}

Let us now consider a pair of entangled universes in the mode $k$ (the index will be omitted for clarity). The density matrix that represents the quantum state of such a pair of universes can be written as
\begin{equation}\label{ISt}
\rho(a) = \mathcal{U}^\dag_S |0_1 0_2\rangle \langle 0_1 0_2 | \mathcal{U}_S ,
\end{equation}
where $|0_1 0_2\rangle \equiv |0_1\rangle |0_2\rangle$, with $|0_1\rangle$ and $|0_2\rangle$ being the ground states of the two entangled universes in their respective Fock spaces, and $\mathcal{U}_S$ is the squeezing evolution operator given by  \cite{RP2012}
\begin{equation}
\mathcal{U}_S(a) = e^{r(a)  \hat{b}_1 \hat{b}_2 - r(a)  \hat{b}_1^\dag \hat{b}_2^\dag} ,
\end{equation}
with $r(a) \equiv {\rm arcsinh} |\nu_p|$, and \cite{Lewis1969, RP2010, RP2012}
\begin{eqnarray}
\mu_p &=& \frac{1}{2\sqrt{\omega(a)}} (\frac{1}{R} + \omega(a) R - i \dot{R}) , \\
\nu_p &=& \frac{1}{2\sqrt{\omega(a)}} (\frac{1}{R} - \omega(a) R - i \dot{R}) .
\end{eqnarray}
The reduced density matrix for each single universe turns out to describe a thermal state  given by \cite{RP2012},
\begin{equation}\label{ThS}
\rho_1(a) = \frac{1}{Z} \sum_{N=0}^\infty e^{-\frac{\omega(a)}{T(a)}(N + \frac{1}{2})} |N\rangle \langle N|,
\end{equation}
where, $|N\rangle \equiv |N\rangle_1$ (similarly for $\rho_2$ with $|N\rangle \equiv |N\rangle_2$), and with,  $Z^{-1} = 2 \sinh\frac{\omega}{2 T}$. The two universes of the entangled pair evolve thus in thermal equilibrium with respect to each other with a temperature,
\begin{equation}\label{eq6327}
T \equiv T(a) = \frac{\omega(a)}{2 \ln\frac{1}{\Gamma(a)}}  ,
\end{equation}
that depends on the scale factor. This kind of thermal states, which have been obtained from an initial entangled pure state, are similar to those consider by Partovi \cite{Partovi2008}, in the context though of a quantum multiverse. Following a parallel reasoning to that made in Refs. \cite{Partovi2008, Jennings2010}, the two thermal state of each single universe, given by Eq. (\ref{ThS}), are locally indistinguishable from classical thermal mixtures. Let us notice that, in the context of the quantum multiverse, by \emph{local} we mean anything that happens in a single universe, i.e. everything that we can observe. Therefore, for any observer the state of the universe appears classical and its state corresponds to a classical mixture. Furthermore, the thermal equilibrium between the two universes of the entangled pair is a stable configuration and any departure from such a configuration turns to return to it \cite{Jennings2010}.

\begin{figure}
\includegraphics[width=8cm,height=6cm]{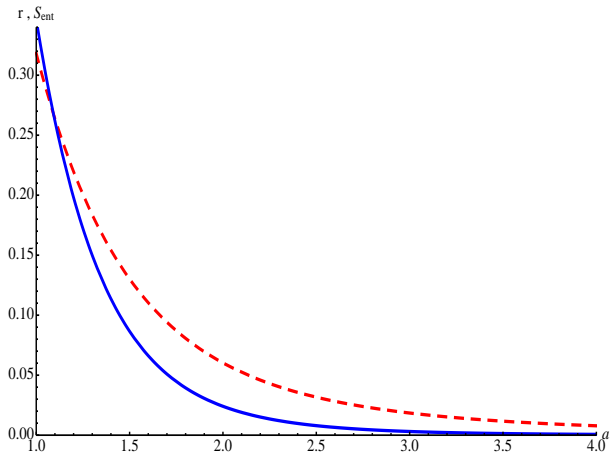}                

\caption{Parameter of squeezing, $r$ (dashed line),  and entropy of entanglement, $S_{ent}$ (continuous line), with respect to the value of the scale factor, $a$.}
\label{fig:entanglement}
\end{figure}

As it is pointed out in Ref. \cite{Jennings2010}, the quantum thermodynamic behaviour of the states given by Eq. (\ref{ThS}) can be quite different from the classical one. However, we shall show that, at least in the model of the multiverse being considered, both the classical and quantum thermodynamical arrows of time agree in requiring the expansion of the single universes of the multiverse once they have crossed the Euclidean quantum barrier.

Let us first notice that the initial state given by Eq. (\ref{ISt}) is a pure state and its entropy is therefore zero. For each single universe, however, there exists a entropy of entanglement given by
\begin{equation}
S_{ent} = - {\rm Tr} (\rho_{1} \ln \rho_{1}) = - {\rm Tr} (\rho_{2} \ln \rho_{2}) ,
\end{equation}
that turns out to be   \cite{RP2012}
\begin{equation}\label{eq6329}
S_{ent}(a) = \cosh^2 r \, \ln \cosh^2 r - \sinh^2 r \, \ln \sinh^2 r .
\end{equation}
It is a growing function with respect to the squeezing parameter, $r$. However, the squeezing parameter and the entanglement rate between the universes decrease with respect to growing values of the scale factor, i.e. $\frac{d S_{ent}}{d a} < 0$ (see Fig. \ref{fig:entanglement}). The second principle of thermodynamics is still satisfied. Here, we must be careful and point out that, unlike it is sometimes stated \cite{Jennings2010}, the second principle of thermodynamics does not state that the entropy cannot ever decrease but that it cannot decrease in any adiabatic process. More specifically, it states that the production of entropy, $\sigma$, defined as \cite{Alicki2004}
\begin{equation}\label{sigma}
\sigma \equiv \frac{d S_{ent}}{da} - \frac{1}{T} \frac{\delta Q}{d a}  ,
\end{equation}
can never be negative ($\sigma \geq 0$). In the quantum multiverse it can be checked, with  \cite{RP2012}
\begin{equation}
\delta Q = \omega \sinh2r dr ,
\end{equation}
that $\sigma = 0$ for any value of the scale factor. Then, the second principle of thermodynamics is strictly satisfied for any expansion or contraction rate of the universe and, thus, it imposes no restriction on the expansion (or contraction) of single universes.

However, the expansion of the universe is followed from the second principle of entanglement thermodynamics \cite{Plenio1998}. In the multiverse, it can be reparaphrased as follows: by local operations and classical communications alone, the amount of entanglement between the universes cannot  increase. Let us recall that by \emph{local} operations we mean in the multiverse anything that happens within a single universe, i.e. everything we can observe. Therefore, the grow of cosmic structures, the particle interactions and even the presence of life in the universe cannot increase the amount of entanglement between the pair of universes provided that all these features are due to local interactions. They should  decrease the rate of entanglement in a non-reversible universe, actually. 

The amount of entanglement between the pair of universes only decreases for growing values of the scale factor (see Fig. \ref{fig:entanglement}). Thus, the second law of the entanglement thermodynamics implies that the universe has to expand once it crosses the quantum barrier at the Euclidean regime, becoming then a large parent universe for which the approximations considered in the model are valid. Furthermore, if the classical thermodynamics and the thermodynamics of entanglement were related, it could be followed that the negative change of entropy as the universe expands favours the creation of cosmic structures and other local processes that increase the local (classical) entropy. The decrease of the entropy of entanglement is larger for a small value of the scale factor. Then, the creation of local structures in the universe would be  favoured in the earliest phases of the universe, as it was expected.

However, the relationship between classical thermodynamics and the thermodynamics of entanglement is not  clear yet. If such a relationship were properly established, then, it could be analyzed in more detail the link between the entropy of entanglement between different universes and the production of local entropy in a single universe.

Let us finally notice that the thermodynamical arrow of time has also implications on the boundary conditions that have to be imposed on single universes. The boundary conditions for the probability amplitudes, $A_k$ in Eq. (\ref{decomposition}), have to be chosen in such a way that they represent expanding branches of the universe. The contracting branches that appear in the linear combination of $B_k(a,\varphi)$, in Eq. (\ref{decomposition2}), would then rapidly disappear as the universe expands and the entanglement rate between universes decreases, where $|\nu|\rightarrow 0$.

\section{Conclusions and further comments}

We have presented the model of a quantum multiverse which is made up of pairs of universes whose quantum mechanical states are entangled with respect to each other within a single pair. The state of the whole multiverse is given by the product of pure states of such entangled pairs of universes. Therefore, the entropy of the whole multiverse is zero. It could be said that \emph{evolution} and \emph{time} are concepts that have  a consistent meaning inside a single universe. Then, it is not expected that the multiverse as a whole posses any arrow of time.

Single universes, however, do have an entropy of entanglement different from zero and  they posse therefore a quantum thermodynamical arrow of time. They are represented by thermal states which are locally indistinguishable from a classical thermal mixture. By local, we mean in the multiverse anything that happens within a single universe, i.e. everything  we can observe. Thus, any observer of the multiverse would see her universe as it were in a classical thermal state. However, the origin of such a  thermal state is essentially of a quantum nature because it has been derived from the entanglement between different universes of the multiverse.

The thermodynamical arrow of time and the arrow of time that is derived from the thermodynamics of entanglement have been studied in the quantum multiverse. The former does not impose any restriction on the expansion of single universes, being such a expansion however demanded by the second principle of the entanglement thermodynamics. Therefore, a single universe should expand once it has become a large universe, far away from the quantum barrier at the Euclidean regime, where the approximations made in the model are valid.

If the thermodynamics of entanglement were properly related to the classical thermodynamics, then, the entanglement arrow of time would favour the grow of local cosmic structures and other physical processes that would increase the amount of local entropy in the earliest stages of a single universe. The thermal state of each single universe is a stable configuration and any departure from it should be small. Thus, the presence of life would be expected in a universe entangled to ours.

However, the model is quite simple to precisely account for the creation of local structures and to describe the  entropy produced by changes in the configuration of the matter fields of the universe. Nevertheless, it seems clear that the multiverse is a rich new paradigm that impels us to look for novel ways of facing up traditional problems in quantum and classical cosmology.

\acknowledgements I would like to thank the participants of the VII Iberian Cosmology Meeting - Ibericos2012 for their interesting and useful comments.

\bibliographystyle{apsrev}
\bibliography{bibliography}

\begin{thebibliography}{37}
\expandafter\ifx\csname natexlab\endcsname\relax\def\natexlab#1{#1}\fi
\expandafter\ifx\csname bibnamefont\endcsname\relax
  \def\bibnamefont#1{#1}\fi
\expandafter\ifx\csname bibfnamefont\endcsname\relax
  \def\bibfnamefont#1{#1}\fi
\expandafter\ifx\csname citenamefont\endcsname\relax
  \def\citenamefont#1{#1}\fi
\expandafter\ifx\csname url\endcsname\relax
  \def\url#1{\texttt{#1}}\fi
\expandafter\ifx\csname urlprefix\endcsname\relax\def\urlprefix{URL }\fi
\providecommand{\bibinfo}[2]{#2}
\providecommand{\eprint}[2][]{\url{#2}}

\bibitem[{\citenamefont{Hawking}(1985)}]{Hawking1985}
\bibinfo{author}{\bibfnamefont{S.~W.} \bibnamefont{Hawking}},
  \bibinfo{journal}{Phys. Rev. D} \textbf{\bibinfo{volume}{32}},
  \bibinfo{pages}{2489} (\bibinfo{year}{1985}).

\bibitem[{\citenamefont{Davies}(1974)}]{Davies1974}
\bibinfo{author}{\bibfnamefont{P.~C.~W.} \bibnamefont{Davies}},
  \emph{\bibinfo{title}{The physics of the time asymmetry}}
  (\bibinfo{publisher}{California University Press, Berkeley},
  \bibinfo{year}{1974}).

\bibitem[{\citenamefont{Zeh}(1989)}]{Zeh1989}
\bibinfo{author}{\bibfnamefont{H.~D.} \bibnamefont{Zeh}},
  \emph{\bibinfo{title}{The physical basis of the direction of time}}
  (\bibinfo{publisher}{Springer-Verlag, Berlin, Germany},
  \bibinfo{year}{1989}).

\bibitem[{\citenamefont{Halliwell et~al.}(1994)\citenamefont{Halliwell,
  P{\'e}rez-Mercader, and Zurek}}]{Halliwell1994}
\bibinfo{editor}{\bibfnamefont{J.~J.} \bibnamefont{Halliwell}},
  \bibinfo{editor}{\bibfnamefont{J.}~\bibnamefont{P{\'e}rez-Mercader}},
  \bibnamefont{and} \bibinfo{editor}{\bibfnamefont{W.~H.} \bibnamefont{Zurek}},
  eds., \emph{\bibinfo{title}{Physical origins of time asymmetry}}
  (\bibinfo{publisher}{Cambridge University Press, Cambridge, UK},
  \bibinfo{year}{1994}).

\bibitem[{\citenamefont{Kiefer and Zeh}(1995)}]{Kiefer1995}
\bibinfo{author}{\bibfnamefont{C.}~\bibnamefont{Kiefer}} \bibnamefont{and}
  \bibinfo{author}{\bibfnamefont{H.~D.} \bibnamefont{Zeh}},
  \bibinfo{journal}{Phys. Rev. D} \textbf{\bibinfo{volume}{51}},
  \bibinfo{pages}{4145} (\bibinfo{year}{1995}), \eprint{gr-qc/9402036v2}.

\bibitem[{\citenamefont{Wald}(2005)}]{Wald2005}
\bibinfo{author}{\bibfnamefont{R.~M.} \bibnamefont{Wald}}
  (\bibinfo{year}{2005}), \eprint{gr-qc/0507094v1}.

\bibitem[{\citenamefont{McInnes}(2007)}]{McInnes2007}
\bibinfo{author}{\bibfnamefont{B.}~\bibnamefont{McInnes}},
  \bibinfo{journal}{Nucl. Phys. B} \textbf{\bibinfo{volume}{782}},
  \bibinfo{pages}{1} (\bibinfo{year}{2007}), \eprint{hep-th/0611088v3}.

\bibitem[{\citenamefont{Gorsky}(2007)}]{Gorsky2007}
\bibinfo{author}{\bibfnamefont{A.}~\bibnamefont{Gorsky}},
  \bibinfo{journal}{Phys. Lett. B} \textbf{\bibinfo{volume}{646}},
  \bibinfo{pages}{183} (\bibinfo{year}{2007}), \eprint{hep-th/0606072}.

\bibitem[{\citenamefont{Bousso}(2011)}]{Bousso2011}
\bibinfo{author}{\bibfnamefont{R.}~\bibnamefont{Bousso}}
  (\bibinfo{year}{2011}), \eprint{arXiv:1112.3341v1}.

\bibitem[{\citenamefont{Goldwirth and Piran}(1991)}]{Goldwirth1991}
\bibinfo{author}{\bibfnamefont{D.~S.} \bibnamefont{Goldwirth}}
  \bibnamefont{and} \bibinfo{author}{\bibfnamefont{T.}~\bibnamefont{Piran}},
  \bibinfo{journal}{Class. Quant. Grav.} \textbf{\bibinfo{volume}{8}},
  \bibinfo{pages}{L155} (\bibinfo{year}{1991}).

\bibitem[{\citenamefont{Albrecht}(2004)}]{Albrecht2003}
\bibinfo{author}{\bibfnamefont{A.}~\bibnamefont{Albrecht}},
  \emph{\bibinfo{title}{Science and ultimate reality. From quantum to cosmos.}}
  (\bibinfo{publisher}{Cambridge University Press, Cambridge, UK},
  \bibinfo{year}{2004}), \eprint{astro-ph/0210527}.

\bibitem[{\citenamefont{Carroll and Chen}(2004)}]{Carroll2004}
\bibinfo{author}{\bibfnamefont{S.~M.} \bibnamefont{Carroll}} \bibnamefont{and}
  \bibinfo{author}{\bibfnamefont{J.}~\bibnamefont{Chen}}
  (\bibinfo{year}{2004}), \eprint{hep-th/0410270}.

\bibitem[{\citenamefont{Gonz{\'a}lez-D{\'\i}az}(1994)}]{PFGD1994}
\bibinfo{author}{\bibfnamefont{P.~F.} \bibnamefont{Gonz{\'a}lez-D{\'\i}az}},
  \bibinfo{journal}{Int. J. Mod. Phys. D} \textbf{\bibinfo{volume}{3}},
  \bibinfo{pages}{549} (\bibinfo{year}{1994}).

\bibitem[{\citenamefont{Castagnino and Laciana}(2002)}]{Castagnino2002}
\bibinfo{author}{\bibfnamefont{M.}~\bibnamefont{Castagnino}} \bibnamefont{and}
  \bibinfo{author}{\bibfnamefont{C.}~\bibnamefont{Laciana}},
  \bibinfo{journal}{Class. Quant. Grav.} \textbf{\bibinfo{volume}{19}},
  \bibinfo{pages}{2657} (\bibinfo{year}{2002}).

\bibitem[{\citenamefont{Kiefer}(2005)}]{Kiefer2005}
\bibinfo{author}{\bibfnamefont{C.}~\bibnamefont{Kiefer}},
  \bibinfo{journal}{Braz. J. Phys.} \textbf{\bibinfo{volume}{35}},
  \bibinfo{pages}{296} (\bibinfo{year}{2005}).

\bibitem[{\citenamefont{Alfinito and Vitiello}(2007)}]{Alfinito2007}
\bibinfo{author}{\bibfnamefont{E.}~\bibnamefont{Alfinito}} \bibnamefont{and}
  \bibinfo{author}{\bibfnamefont{G.}~\bibnamefont{Vitiello}},
  \bibinfo{journal}{J. Phys.: Conf. Ser.} \textbf{\bibinfo{volume}{67}},
  \bibinfo{pages}{012010} (\bibinfo{year}{2007}).

\bibitem[{\citenamefont{Bertolami}(2008)}]{Bertolami2008}
\bibinfo{author}{\bibfnamefont{O.}~\bibnamefont{Bertolami}},
  \bibinfo{journal}{Gen. Re. Grav.} \textbf{\bibinfo{volume}{40}},
  \bibinfo{pages}{1891} (\bibinfo{year}{2008}), \eprint{arXiv:0705.2325v1}.

\bibitem[{\citenamefont{Carr et~al.}(2007)}]{Carr2007}
\bibinfo{editor}{\bibfnamefont{B.}~\bibnamefont{Carr}} \bibnamefont{et~al.},
  eds., \emph{\bibinfo{title}{Universe or Multiverse}}
  (\bibinfo{publisher}{Cambridge University Press, Cambridge, UK},
  \bibinfo{year}{2007}).

\bibitem[{\citenamefont{Gonz{\'a}lez-D{\'\i}az}()}]{PFGD2012}
\bibinfo{author}{\bibfnamefont{P.~F.} \bibnamefont{Gonz{\'a}lez-D{\'\i}az}},
  \bibinfo{journal}{(In preparation)}  (????).

\bibitem[{\citenamefont{Robles-P{\'e}rez and
  Gonz{\'a}lez-D{\'\i}az}(2010)}]{RP2010}
\bibinfo{author}{\bibfnamefont{S.}~\bibnamefont{Robles-P{\'e}rez}}
  \bibnamefont{and} \bibinfo{author}{\bibfnamefont{P.~F.}
  \bibnamefont{Gonz{\'a}lez-D{\'\i}az}}, \bibinfo{journal}{Phys. Rev. D}
  \textbf{\bibinfo{volume}{81}}, \bibinfo{pages}{083529}
  (\bibinfo{year}{2010}), \eprint{arXiv:1005.2147v1}.

\bibitem[{\citenamefont{Robles-P{\'e}rez and
  Gonz{\'a}lez-D{\'\i}az}(2012)}]{RP2012}
\bibinfo{author}{\bibfnamefont{S.}~\bibnamefont{Robles-P{\'e}rez}}
  \bibnamefont{and} \bibinfo{author}{\bibfnamefont{P.~F.}
  \bibnamefont{Gonz{\'a}lez-D{\'\i}az}}, \bibinfo{journal}{(submitted to PRD
  for publication)}  (\bibinfo{year}{2012}), \eprint{1111.4128v1}.

\bibitem[{\citenamefont{Jennings and Rudolph}(2010)}]{Jennings2010}
\bibinfo{author}{\bibfnamefont{D.}~\bibnamefont{Jennings}} \bibnamefont{and}
  \bibinfo{author}{\bibfnamefont{T.}~\bibnamefont{Rudolph}},
  \bibinfo{journal}{Phys. Rev. E} \textbf{\bibinfo{volume}{81}},
  \bibinfo{pages}{061130} (\bibinfo{year}{2010}).

\bibitem[{\citenamefont{Vedral and Kashefi}(2002)}]{Vedral2002}
\bibinfo{author}{\bibfnamefont{V.}~\bibnamefont{Vedral}} \bibnamefont{and}
  \bibinfo{author}{\bibfnamefont{E.}~\bibnamefont{Kashefi}},
  \bibinfo{journal}{Phys. Rev. Lett.} \textbf{\bibinfo{volume}{89}},
  \bibinfo{pages}{3} (\bibinfo{year}{2002}).

\bibitem[{\citenamefont{Brandao and Plenio}(2008)}]{Brandao2008}
\bibinfo{author}{\bibfnamefont{F.~G. S.~L.} \bibnamefont{Brandao}}
  \bibnamefont{and} \bibinfo{author}{\bibfnamefont{M.~B.}
  \bibnamefont{Plenio}}, \bibinfo{journal}{Nature Physics}
  \textbf{\bibinfo{volume}{4}}, \bibinfo{pages}{873} (\bibinfo{year}{2008}).

\bibitem[{\citenamefont{Deutsch}(2002)}]{Deutsch2002}
\bibinfo{author}{\bibfnamefont{D.}~\bibnamefont{Deutsch}}, in
  \emph{\bibinfo{booktitle}{Proc. R. Soc. Lond. A}} (\bibinfo{year}{2002}),
  vol. \bibinfo{volume}{458}, pp. \bibinfo{pages}{2911--2923},
  \eprint{quant-ph/0104033}.

\bibitem[{\citenamefont{Strominger}(1990)}]{Strominger1990}
\bibinfo{author}{\bibfnamefont{A.}~\bibnamefont{Strominger}}, in
  \emph{\bibinfo{booktitle}{Quantum Cosmology and Baby Universes}}, edited by
  \bibinfo{editor}{\bibfnamefont{S.}~\bibnamefont{Coleman}},
  \bibinfo{editor}{\bibfnamefont{J.~B.} \bibnamefont{Hartle}},
  \bibinfo{editor}{\bibfnamefont{T.}~\bibnamefont{Piran}}, \bibnamefont{and}
  \bibinfo{editor}{\bibfnamefont{S.}~\bibnamefont{Weinberg}}
  (\bibinfo{publisher}{World Scientific, London, UK}, \bibinfo{year}{1990}),
  vol.~\bibinfo{volume}{7}.

\bibitem[{\citenamefont{Litt et~al.}(2006)}]{Litt2006}
\bibinfo{author}{\bibfnamefont{A.}~\bibnamefont{Litt}} \bibnamefont{et~al.},
  \bibinfo{journal}{Cognitive Science} \textbf{\bibinfo{volume}{30}},
  \bibinfo{pages}{593} (\bibinfo{year}{2006}).

\bibitem[{\citenamefont{Kiefer}(2007)}]{Kiefer2007}
\bibinfo{author}{\bibfnamefont{C.}~\bibnamefont{Kiefer}},
  \emph{\bibinfo{title}{Quantum gravity}} (\bibinfo{publisher}{Oxford
  University Press, Oxford, UK}, \bibinfo{year}{2007}).

\bibitem[{\citenamefont{Vilenkin}(1982)}]{Vilenkin1982}
\bibinfo{author}{\bibfnamefont{A.}~\bibnamefont{Vilenkin}},
  \bibinfo{journal}{Phys. Lett. B} \textbf{\bibinfo{volume}{117}},
  \bibinfo{pages}{25} (\bibinfo{year}{1982}).

\bibitem[{\citenamefont{Vilenkin}(1984)}]{Vilenkin1984}
\bibinfo{author}{\bibfnamefont{A.}~\bibnamefont{Vilenkin}},
  \bibinfo{journal}{Phys. Rev. D} \textbf{\bibinfo{volume}{30}},
  \bibinfo{pages}{509} (\bibinfo{year}{1984}).

\bibitem[{\citenamefont{Vilenkin}(1986)}]{Vilenkin1986}
\bibinfo{author}{\bibfnamefont{A.}~\bibnamefont{Vilenkin}},
  \bibinfo{journal}{Phys. Rev. D} \textbf{\bibinfo{volume}{33}},
  \bibinfo{pages}{3560} (\bibinfo{year}{1986}).

\bibitem[{\citenamefont{Hawking}(1982)}]{Hawking1982}
\bibinfo{author}{\bibfnamefont{S.~W.} \bibnamefont{Hawking}},
  \bibinfo{journal}{Astrophysical Cosmology, 563-72. Vatican City: Pontificia
  Academiae Scientarium}  (\bibinfo{year}{1982}).

\bibitem[{\citenamefont{Hartle and Hawking}(1983)}]{Hartle1983}
\bibinfo{author}{\bibfnamefont{J.~B.} \bibnamefont{Hartle}} \bibnamefont{and}
  \bibinfo{author}{\bibfnamefont{S.~W.} \bibnamefont{Hawking}},
  \bibinfo{journal}{Phys. Rev. D} \textbf{\bibinfo{volume}{28}},
  \bibinfo{pages}{2960} (\bibinfo{year}{1983}).

\bibitem[{\citenamefont{Lewis and Riesenfeld}(1969)}]{Lewis1969}
\bibinfo{author}{\bibfnamefont{H.~R.} \bibnamefont{Lewis}} \bibnamefont{and}
  \bibinfo{author}{\bibfnamefont{W.~B.} \bibnamefont{Riesenfeld}},
  \bibinfo{journal}{J. Math. Phys.} \textbf{\bibinfo{volume}{10}},
  \bibinfo{pages}{1458} (\bibinfo{year}{1969}).

\bibitem[{\citenamefont{Partovi}(2008)}]{Partovi2008}
\bibinfo{author}{\bibfnamefont{M.~H.} \bibnamefont{Partovi}},
  \bibinfo{journal}{Phys. Rev. E} \textbf{\bibinfo{volume}{77}},
  \bibinfo{pages}{021110} (\bibinfo{year}{2008}).

\bibitem[{\citenamefont{Alicki et~al.}(2004)}]{Alicki2004}
\bibinfo{author}{\bibfnamefont{R.}~\bibnamefont{Alicki}} \bibnamefont{et~al.},
  \bibinfo{journal}{Open Syst. Inf. Dyn.} \textbf{\bibinfo{volume}{11}},
  \bibinfo{pages}{205} (\bibinfo{year}{2004}).

\bibitem[{\citenamefont{Plenio and Vedral}(1998)}]{Plenio1998}
\bibinfo{author}{\bibfnamefont{M.~B.} \bibnamefont{Plenio}} \bibnamefont{and}
  \bibinfo{author}{\bibfnamefont{V.}~\bibnamefont{Vedral}},
  \bibinfo{journal}{Comtemp. Phys.} \textbf{\bibinfo{volume}{39}},
  \bibinfo{pages}{431} (\bibinfo{year}{1998}).

\end{thebibliography}

\end{document}